\documentclass[aps,prl,twocolumn,groupedaddress,showpacs,letterpaper]{revtex4}
\usepackage{amsmath,amssymb,graphicx,hyperref}
\hypersetup{pdfnewwindow=true, colorlinks=true, linkcolor=blue, anchorcolor=blue,
  citecolor=blue, filecolor=blue, menucolor=blue, urlcolor=blue}

\def\ket#1{\vert #1 \rangle}
\def\bra#1{\langle #1 \vert}
\def\me#1#2#3{\bra{#1} #2 \ket{#3}}

\def\avg#1{\langle #1 \rangle}
\DeclareMathOperator{\sgn}{sgn}

\begin{document}

\title{Valence Fluctuations and Quasiparticle Multiplets in Pu Chalcogenides and Pnictides}

\author{Chuck-Hou Yee}
\email{chuckyee@physics.rutgers.edu}
\author{Gabriel Kotliar}
\author{Kristjan Haule}
\affiliation{Department of Physics \& Astronomy, Rutgers University, Piscataway, NJ 08854-8019, USA}

\date{\today}

\begin{abstract}
The spectra of Pu chalcogenides and pnictides are computed with LDA+DMFT and interpreted with the aid of valence histograms and slave-boson calculations.  We find the chalcogenides are mixed-valent ($n_f = 5.2$) materials with a strongly $T$-dependent low-energy density of states and a triplet of quasiparticle peaks below the Fermi level.  Furthermore, we predict a doublet of reflected peaks above the Fermi level.  In the pnictides, the raising of $f^6$ states relative to $f^5$ suppresses valence fluctuations, resulting in integral-valent ($n_f = 5.0$) local moment metals.
\end{abstract}

\pacs{71.27.+a, 75.30.Mb}

\maketitle

The stark contrast in behavior between the plutonium monochalcogenides and monopnictides is a longstanding issue in strongly-correlated physics.  The pnictides (PuSb, PuAs, PuP) are comparatively simple metals~\cite{Blaise85} with embedded $f$-moments arising from trivalent Pu ions which order in the range $\Theta_\text{CW} = 85$ to $126$~K~\cite{Lander87}.  In contrast, the chalcogenides (PuTe, PuSe, PuS) exhibit seemingly contradictory behavior: they have a large room temperature specific heat~\cite{Stewart91}, yet the resistivity indicates a small gap~\cite{Therond87,Fournier90,Ichas01}.  Also, their lattice constant rules out the full-shell divalent Pu state, yet the susceptibility shows no evidence of Curie-Weiss behavior~\cite{Lander87}.  Furthermore, photoemission observes a triplet of peaks (the ``photoemission triplet'') near the Fermi level~\cite{Gouder00,Havela02,Wachter03,Durakiewicz04} whose origin is still hotly debated~\cite{Gouder00,Havela02}.

The contrast between the Pu chalcogenides and pnictides exemplifies the view that the Pu $5f$ electrons sit at the edge of a localization-delocalization transition, where small changes in their electronic environment can drive a transition to itinerancy or localization, thus posing a major challenge to electronic structure methods.  Theoretical studies of the chalcogenides within LDA~\cite{Oppeneer00,Shorikov05} predict a metal and do not account for the photoemission triplet.  Methods treating correlations beyond LDA have improved the situation, but cannot fully integrate the available experimental data within a single theory.  Non-charge-self-consistent LDA+DMFT with FLEX predicts metallic behavior and misses the photoemission triplet~\cite{Pourovskii05}.  Adding charge-self-consistency~\cite{Suzuki09} opens a gap, but still misses the photoemission triplet.  LDA+DMFT with either ED in a small Hilbert space~\cite{Svane06} or Hubbard-I~\cite{Shick07} as the impurity solver describes the photoemission triplet but fails to explain the resistivity.

\begin{figure}[b]
\includegraphics[width=\linewidth]{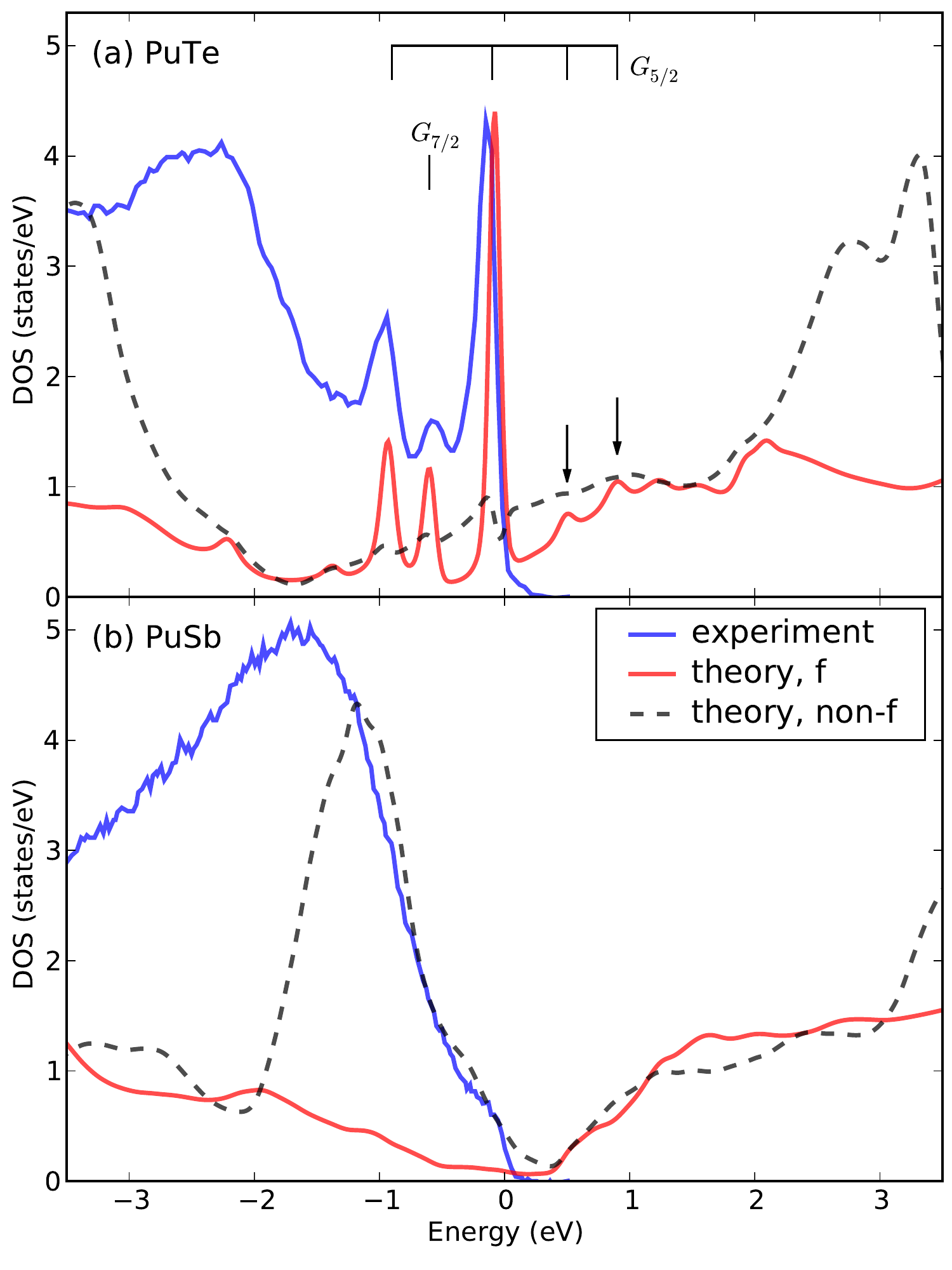}
\caption{Computed spectrum for (a) PuTe and (b) PuSb compared with photoemission~\cite{Durakiewicz04}.  A triplet of peaks are present in PuTe, as well as a predicted reflected doublet of peaks (arrows), while neither appear in PuSb.  Application of broadening ($40$~meV) has blurred the gap in PuTe (see Fig. \ref{gap}).} \label{dos}
\end{figure}

In this Letter, we elucidate the mechanism driving the electronic trends between the pnictides and chalcogenides within a single framework.  We find that the chalcogenides are mixed-valent compounds where valence fluctuations combine with the underlying Pu atomic multiplet structure to drive the formation of a multiplet of many-body quasiparticle peaks (``quasiparticle multiplets'') which correspond to the observed photoemission triplet.  These heavy quasiparticles strongly affect the density of states at the Fermi level as a function of temperature, corroborating the gap-like resistivity and large specific heat at room temperature.  Using analytic methods, we provide a description of the quasiparticle multiplet formation and their coexistence with the development of a gap.  In contrast, the chemistry of the pnictides shifts the atomic multiplet energies, rendering valence fluctuations too costly, thereby localizing the $f$ electrons.

We use LDA+DMFT~\cite{KotliarRMP,Held} with OCA as the impurity solver to model the chalcogenides and pnictides, taking PuTe and PuSb as representatives of the two groups due to the special attention~\cite{Therond87,Ichas01,Gouder00,Durakiewicz04,Blaise85} accorded to them in the available experimental data.  In our calculations, we use the projective orthogonalized LMTO basis set~\cite{Toropova07}.  We use $U=4.5$~eV for the Coulomb interaction, consistent with previous work~\cite{Savrasov01,Zhu07,Shim07}. The Slater integrals $F^2 = 6.1$~eV, $F^4=4.1$~eV and $F^6=3.0$~eV are calculated using Cowan's atomic structure code~\cite{Cowan} and reduced by $30$~\% to account for screening.

\begin{figure}
\includegraphics[width=\linewidth]{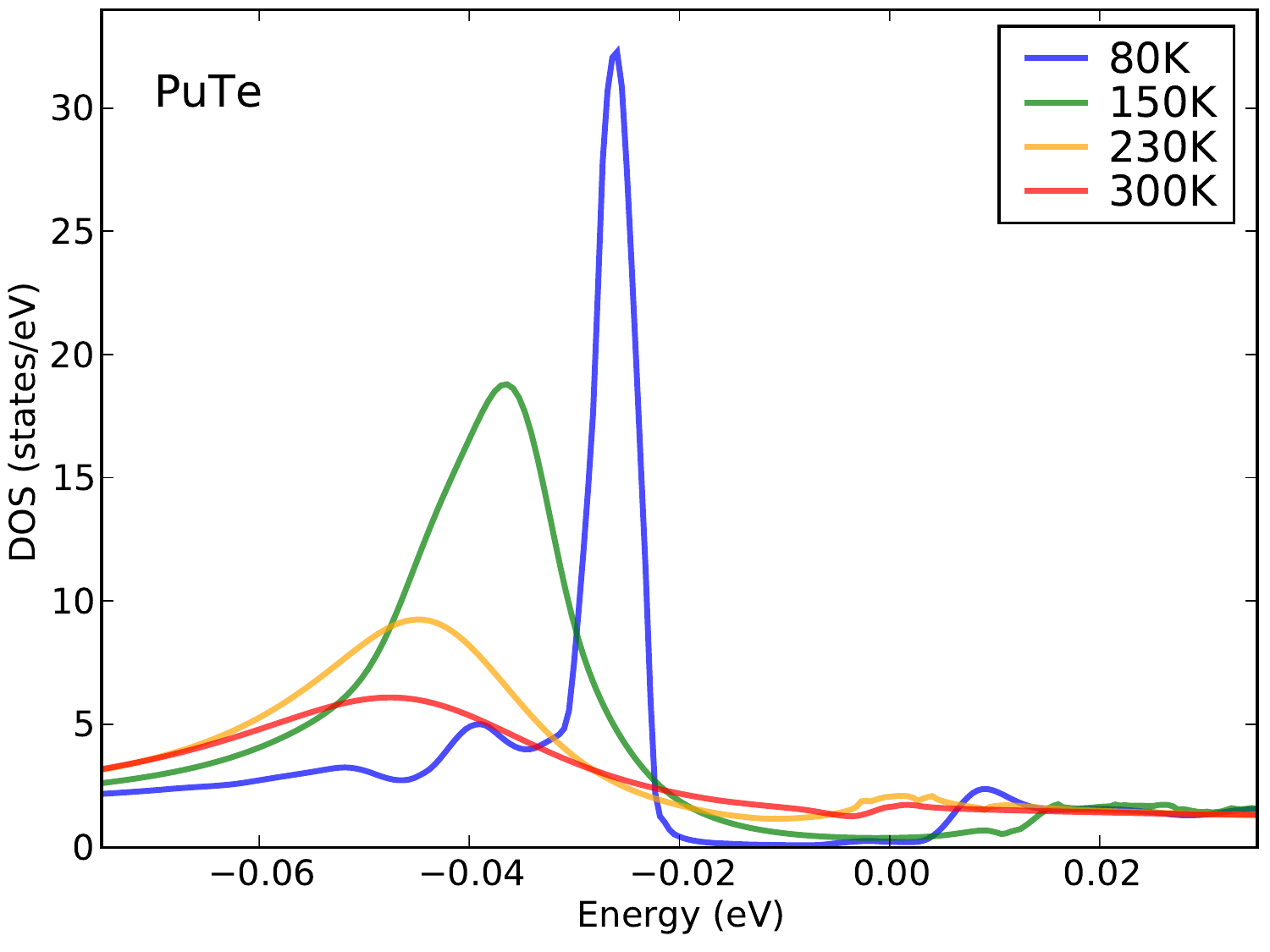}
\caption{Detail of PuTe spectrum near Fermi level, showing development of gap and formation of main quasiparticle peak with decreasing temperature.} \label{gap}
\end{figure}

In Fig. \ref{dos}, we show the computed spectral functions for PuTe and PuSb, resolution-broadened by $40$~meV and overlaid with experimental photoemission data~\cite{Durakiewicz04}.  The calculations clearly corroborate the presence of the photoemission triplet in PuTe at the correct energies, and their absence in PuSb.  Furthermore, we predict the existence of a doublet of peaks (arrows) in PuTe at reflected energies about the Fermi level.  The strong temperature dependence of all five peaks indicates they are quasiparticle resonances.  Examining the main quasiparticle peak at the Fermi level (Fig. \ref{gap}), we find it is composed of heavily renormalized quasiparticles with $Z \approx 0.1$, giving a greatly enhanced specific heat.  Additionally, the peak sharpens with decreasing temperature, considerably reducing the density of states at the Fermi level, leading to the formation of a gap and the observed temperature dependence in the specific heat~\cite{Stewart91}.  Together, the reduction of Fermi level density and heavy renormalization explain how a gap-like resistivity can coexist with a large specific heat coefficient at room temperatures.

\begin{figure}
\includegraphics[width=\linewidth]{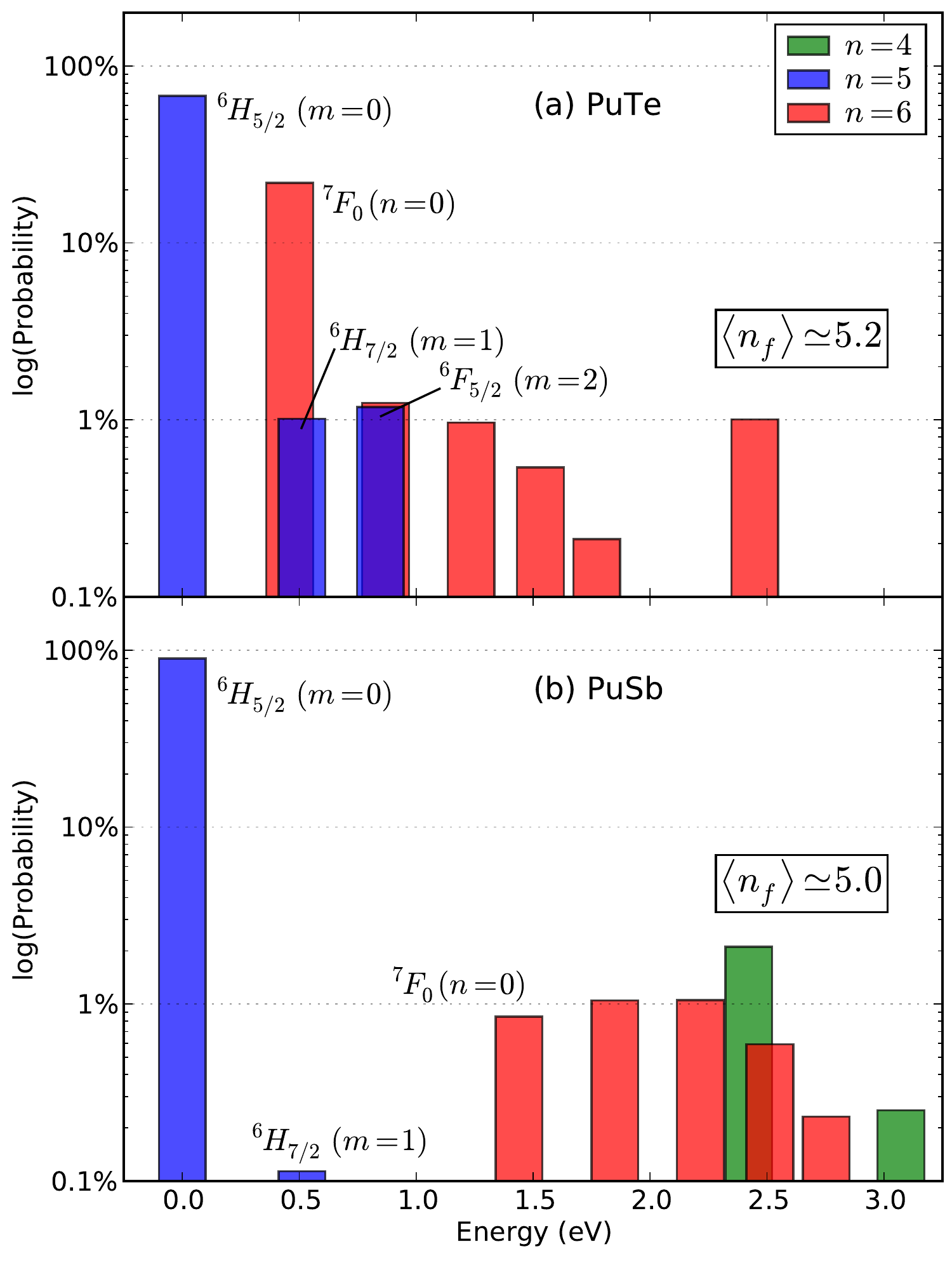}
\caption{Valence histograms obtained by projection of DMFT solution ($T=60$~K) onto Pu atomic eigenstates, plotted with energies relative to lowest-energy atomic state.  The height of each bar represents the percentage of time the atom spends in each configuration.  PuTe is strongly mixed-valent due to the small energy cost ($\approx 0.5$~eV) of valence fluctuations from $f^5$ to $f^6$, while PuSb is integral valent due to the large cost of fluctuations to both $f^4$ and $f^6$.  In PuTe, additional fluctuations to atomic multiplets ${}^6H_{7/2}$ and ${}^6F_{5/2}$ are crucial to the creation of the photoemission triplet.  All quantum numbers are gathered into a single index $m$ or $n$ used in the slave-boson calculation.  Approximate term symbols are given although $L$ and $S$ strictly are not good quantum numbers.} \label{histo}
\end{figure}

A useful way to analyze the Pu atomic environment is to quantify the amount of time the $f$-electrons spend in each atomic configuration as they fluctuate between the atom and conduction band.  To this end, we project the DMFT ground state $\ket{\Omega}$ onto the Pu $f$-electron atomic eigenstates, resulting in the probabilities $P_m = Z^{-1}\me{\Omega}{X_{mm}}{\Omega}$, where $Z = \sum_m \me{\Omega}{X_{mm}}{\Omega}$ is the normalization and $X_{mm}$ is the Hubbard operator which projects onto the $m$th atomic eigenstate~\cite{Shim07}.  Plotting $P_m$ against the atomic energies gives a valence histogram (Fig. \ref{histo}) which graphically represent the relative weights of the atomic configurations comprising $\ket{\Omega}$.  The $f$ valence can then be defined by $\avg{n_f} = \sum_m P_m n_m$, where $n_m$ is the number of electrons in the $m$th state.

The histograms show that the Pu atom is restricted to just one or two valences in both compounds.  In fact, the atom mostly exists in a single $f^5$ configuration (${}^6 H_{5/2}$, tall blue bar), which we loosely call the ``ground state''.  However, PuTe differs from PuSb in that its electrons have an over $20$~\% probability to fluctuate to the lowest $f^6$ state (${}^7F_0$, leftmost red bar) due to this state's small $0.5$~eV separation from the ``ground state''.  The resulting mixed-valent ($\avg{n_f} = 5.2$) Pu atom strongly suggests a Kondo-like, and thus nonmagnetic, ground state in PuTe.  Additionally, the proximity in energy of the next two higher $f^5$ states (${}^6H_{7/2}$ at $0.5$~eV and ${}^6F_{5/2}$ at $0.9$~eV) renders these multiplets accessible to valence fluctuations, which will play a role in generating the photoemission triplet.  In contrast, the Pu atom is integral-valent ($\avg{n_f} = 5.0$) in PuSb.  Within LDA, the $j=5/2$ $f$-bands are higher in energy in PuSb, raising the energy of the $f^6$ states relative to the $f^5$.  The resulting $1.5$~eV gap locks Pu into the lowest $f^5$ state and PuSb remains a local moment metal.

To gain additional insight into the LDA+DMFT solution, we construct a Hamiltonian for the DMFT quantum impurity.  The histograms indicate we only need to keep two valences in the atomic Hilbert space for a low-energy model,
\begin{equation}
  H_\text{atom} = \sum_m E^f_m f_m^\dagger f_m + \sum_n E^b_n b_n^\dagger b_n,
\end{equation}
where the auxiliary fermions $f_m^\dagger\ket{0} = \ket{m;f^5}$ and bosons $b_n^\dagger\ket{0} = \ket{n;f^6}$ create the atomic eigenstates, and $E^f_m$ and $E^b_{n}$ are the corresponding atomic eigenenergies.  The Hamiltonian is supplemented by the constraint $Q = \sum_m f_m^\dagger f_m + \sum_n b_n^\dagger b_n = 1$, in the same spirit as the slave-boson construction~\cite{Coleman84,Kotliar86}.  The atom hybridizes with an auxiliary conduction bath,
\begin{equation}
  H_\text{c,mix} = \sum_{k\alpha}\epsilon_{k\alpha} n_{k\alpha} + \sum_{k\alpha}(V_{k\alpha}d^\dagger_\alpha c_{k\alpha} + \text{h.c.}),
\end{equation}
where $d^\dagger_\alpha$ creates an electron in the $\alpha$th atomic crystal field basis and $k$ is the dispersion of the conduction bath.  Since we work in the atomic eigenbasis, we eliminate $d^\dagger$ in favor of the auxiliary particles by expanding $d^\dagger_\alpha = b_n^\dagger (F^{\alpha\dagger})_{nm}f_m$, where $(F^{\alpha\dagger})_{nm} = \me{n}{d_\alpha^\dagger}{m}$ are the matrix elements of the physical electron creation operator.

This model is equivalent to the slave-boson treatment of the multi-orbital Anderson impurity model~\cite{Bickers87,Kroha03}, so we can compute the mean-field solution and fluctuations.  At the mean-field level, we replace the bosonic operators by their averages, $\avg{b_n^\dagger}^2 = \avg{b_n}^2 \equiv z_n$ which are the probabilities of the $f^6$ atomic states (red bars in Fig. \ref{histo}).  Then, the physical Green's function is
\begin{equation}
  G_{\alpha'\alpha}(i\omega) = \!\!\!\!\sum_{m'n'nm}\!\!\!\! F^{\alpha'}_{m'n'} (F^{\alpha\dagger})_{nm} \sqrt{z_{n'}z_n} G^f_{m'm}(-i\omega), \label{Gd}
\end{equation}
where the auxiliary $f$ propagator and hybridization are
\begin{align}
  G^f(i\omega)_{m'm}^{-1} &= (i\omega-E^f_m-\lambda)\delta_{m'm}+i\Delta_{m'm}\sgn\omega, \\
  \Delta_{m'm} &= \sum_{n'n\alpha} F^\alpha_{m'n'} (F^{\alpha\dagger})_{nm} \sqrt{z_{n'}z_n}\Delta_{\alpha\alpha},
\end{align}
and the hybridization $\Delta_{\alpha\alpha'}$ is approximated as an energy-independent constant.  Here, $\lambda$ is the Lagrange multiplier used to maintain $\avg{Q}=1$.  The crucial minus sign $G_{m'm}^f(-i\omega)$ arises because the propagation of a physical electron $\sim\avg{d(\tau)d^\dagger}$ corresponds to an $f$-hole.

In the Kondo regime, the mean-field equations give
\begin{equation}
  T_\text{K} \simeq D e^{-\frac{\pi(\avg{E^b}-E^f_0)}{\Delta_{00}/z}} \prod_{m\neq 0} \left(\frac{D}{E^f_m-E^f_0}\right)^{\Delta_{mm}/\Delta_{00}}, \label{TK}
\end{equation}
where $z = \sum_n z_n$ is the total $f^6$ probability and $\avg{E^b} = z^{-1}\sum_n E^b_n z_n$ is the weighted average of the $f^6$ energy levels.  We ignored the off-diagonal components of $G^f_{m'm}$, which are negligible compared to the diagonal components when $|E^f_{m'} - E^f_m| \gg \Delta_{m'm}$.  In the Kondo regime, $\lambda \approx -E_0^f$, pinning the lowest $f$ propagator near the Fermi level.

For PuTe, we explicitly evaluate the sum in (\ref{Gd}) to determine the origin of the photoemission triplet.  Keeping just $z_0 \approx 0.20$ since the remaining $z_n$ are negligible (see Fig.~\ref{histo}), we find that only three matrix elements have significant weight: $F^{5/2}_{00}$, $F^{5/2}_{02}$ and $F^{7/2}_{01}$.  Again ignoring off-diagonal terms, we find for the physical Green's function
\begin{align}
  G_{5/2}(\omega) &= z_0 |F^{5/2}_{00}|^2 G^f_{00}(-\omega) + z_0 |F^{5/2}_{02}|^2 G^f_{22}(-\omega), \\
  G_{7/2}(\omega) &= z_0 |F^{7/2}_{01}|^2 G^f_{11}(-\omega).
\end{align}
The selection rules contained in the matrix elements determine the spin-orbit structure of the spectrum: $G_{5/2}$ contains the main peak at the Fermi level and the weaker peak at $-0.9$~eV (labeled by brackets in Fig. \ref{dos}), while $G_{7/2}$ contributes the third peak at $-0.5$~eV.  Plugging in $D=1.0$~eV and $\Delta_{5/2} = \Delta_{7/2} = 0.04$~eV into (\ref{TK}) gives $T_\text{K} \approx 500$~K in PuTe, so the peaks have sufficient width to be seen in photoemission.  Numerical solution of the mean-field equations with a small hybridization gap, $\Delta_{\alpha\alpha}(\omega) = \Delta_{\alpha\alpha}[\theta(\omega-E_g)+\theta(-\omega-E_g)]$, confirms that for $2E_g \lesssim T_\text{K}$, the resonances are not destroyed.  Thus, in PuTe, valence fluctuations create three Kondo peaks in the DMFT quantum impurity with spacings determined by the underlying atomic multiplets, corresponding to the quasiparticle triplet in the lattice.

Proceeding to PuSb, the decrement in valence allows Pu to fully transfer three electrons to the pnictogen and exist purely in a trivalent state.  Energetically, this is accomplished by raising the $f^6$ multiplets with respect to the $f^5$ states, inducing a four-fold increase in $\avg{E^b}-E^f_0$ from $0.5$~eV to $2.0$~eV.  This exponentially suppresses the Kondo temperature of PuSb to under $1$~K, eliminating the Kondo peaks, localizing the $f$-electrons and allowing magnetically-ordered states at low temperature.

\begin{figure}
\includegraphics[width=\linewidth]{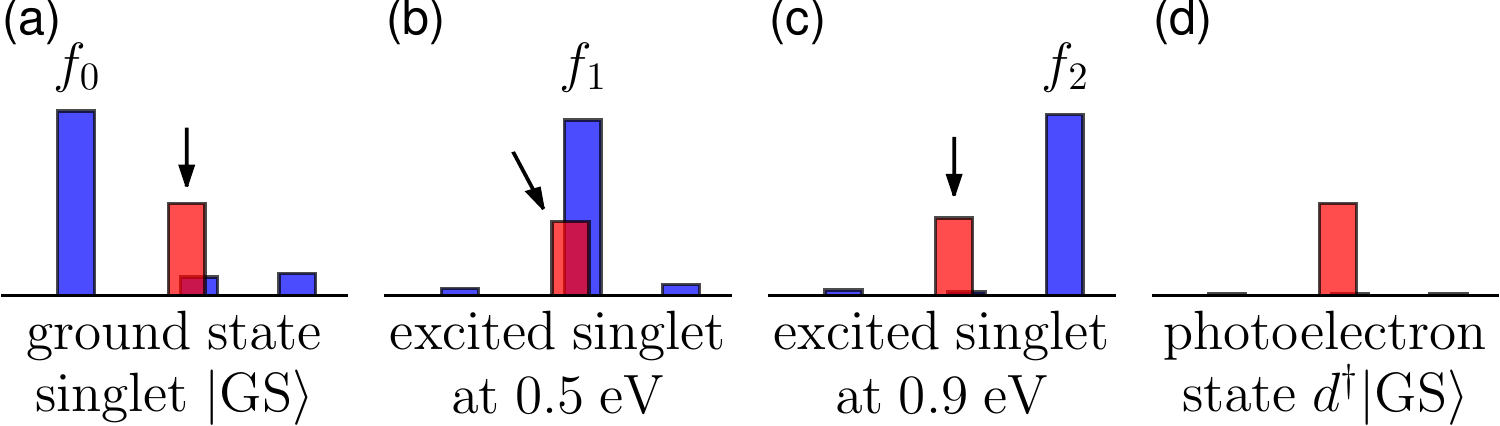}
\caption{Valence histograms of states responsible for the predicted doublet of peaks above the Fermi level in PuTe.  Histograms (a)-(c) depict the ground state and two excited Kondo singlets respectively.  All three state have significant $f^6$ weight (arrows) due to strong valence fluctuations.  Histogram (d) is the photoelectron state $d^\dagger\ket{\text{GS}}$ generated by the addition of an electron to the ground state.  Since $d^\dagger$ destroys $f_m$ and creates $b^\dagger$, the result is a state which is entirely $f^6$ in character.  The large $f^6$ admixture in the excited singlets means they will overlap strongly with the photoelectron state, giving rise to the doublet of peaks at $0.5$~eV and $0.9$~eV.}
\label{histos}
\end{figure}

To explain the Hubbard bands and reflected doublet of peaks above the Fermi level, we compute corrections to the mean-field solution.  These corrections show that the bare atomic multiplets generate Hubbard bands, which are too broad to be seen in Fig. \ref{dos}.  We emphasize that the photoemission triplet is not directly attributed to atomic multiplets, but rather to quasiparticles.  Additionally, the corrections to mean-field show that the reflected doublet of peaks arises from overlap of the ground state Kondo singlet with two excited Kondo singlets (Fig. \ref{histos}).  While the ground state is primarily a singlet formed between $f_0$ and a conduction electron, the two excited states are singlets formed with $f_1$ and $f_2$ in place of $f_0$.  Since the atomic multiplets $f_1$ and $f_2$ lie at energies $0.5$~eV and $0.9$~eV above $f_0$, the two excited Kondo singlets lie at these energies as well.  To generate the spectrum, an electron is first added to the ground state singlet to create a photoelectron state (Fig. \ref{histos}d), which is overlapped with the excited singlets.  The overlap is non-zero since all states have an $f^6$ component due to valence fluctuations, resulting in a doublet of reflected peaks.

To summarize, our LDA+DMFT calculations account for the complex trends observed in experiment across the Pu chalcogenides and pnictides.  These include the formation of a low-temperature gap at the Fermi level accompanied by quasiparticle multiplets in the photoemission spectrum.  Our theory elucidates the mechanism for the emergence of the quasiparticle multiplet excitations.  These excitations represent the remnants of the atomic structure in the low-energy spectra, which is entirely described in terms of quasiparticles.  The spectra and $f$-occupancy confirm that the chalcogenides are correlated low-carrier materials in the mixed-valent regime. The chemistry of the pnictides increases the cost of fluctuations and renders the pnictides divalent, thereby eliminating the quasiparticle multiplets.

Our theory has several experimental consequences for the chalcogenides: there should be a quasiparticle doublet at positive energies $0.5$~eV and $0.9$~eV which can be probed by inverse photoemission techniques.  Furthermore, the quasiparticle multiplets can be detected as side-peaks in the optical conductivity, again at $0.5$~eV and $0.9$~eV.  It would be interesting to study the temperature dependence of both the photoemission and optics near the coherence temperature ($500$~K) as we expect strong temperature dependence due to the many-body nature of the quasiparticle multiplets.  Finally, the mechanism we outline is fairly general and applies to other correlated materials, provided the coherence temperature is large enough to be observed and less than the atomic multiplet splitting.

C.Y. acknowledges useful discussions with P.~Coleman.  This work was supported by NSF Grant DMR-0806937 and BES-DOE Grant DE-FG02-99ER45761.

\bibliography{pux}

\begin{thebibliography}{27}
\expandafter\ifx\csname natexlab\endcsname\relax\def\natexlab#1{#1}\fi
\expandafter\ifx\csname bibnamefont\endcsname\relax
  \def\bibnamefont#1{#1}\fi
\expandafter\ifx\csname bibfnamefont\endcsname\relax
  \def\bibfnamefont#1{#1}\fi
\expandafter\ifx\csname citenamefont\endcsname\relax
  \def\citenamefont#1{#1}\fi
\expandafter\ifx\csname url\endcsname\relax
  \def\url#1{\texttt{#1}}\fi
\expandafter\ifx\csname urlprefix\endcsname\relax\def\urlprefix{URL }\fi
\providecommand{\bibinfo}[2]{#2}
\providecommand{\eprint}[2][]{\url{#2}}

\bibitem[{\citenamefont{Blaise et~al.}(1985)}]{Blaise85}
\bibinfo{author}{\bibfnamefont{A.}~\bibnamefont{Blaise}} \bibnamefont{et~al.},
  \bibinfo{journal}{Physica B+C} \textbf{\bibinfo{volume}{130}},
  \bibinfo{pages}{99 } (\bibinfo{year}{1985}).

\bibitem[{\citenamefont{Lander et~al.}(1987)}]{Lander87}
\bibinfo{author}{\bibfnamefont{G.~H.} \bibnamefont{Lander}}
  \bibnamefont{et~al.}, \bibinfo{journal}{Physica B+C}
  \textbf{\bibinfo{volume}{146}}, \bibinfo{pages}{341 } (\bibinfo{year}{1987}).

\bibitem[{\citenamefont{Stewart et~al.}(1991)}]{Stewart91}
\bibinfo{author}{\bibfnamefont{G.}~\bibnamefont{Stewart}} \bibnamefont{et~al.},
  \bibinfo{journal}{J. Alloys Compnds} \textbf{\bibinfo{volume}{177}},
  \bibinfo{pages}{167 } (\bibinfo{year}{1991}).

\bibitem[{\citenamefont{Therond et~al.}(1987)}]{Therond87}
\bibinfo{author}{\bibfnamefont{P.~G.} \bibnamefont{Therond}}
  \bibnamefont{et~al.}, \bibinfo{journal}{J. Magn. Magn. Mat.}
  \textbf{\bibinfo{volume}{63-64}}, \bibinfo{pages}{142 }
  (\bibinfo{year}{1987}).

\bibitem[{\citenamefont{Fournier et~al.}(1990)}]{Fournier90}
\bibinfo{author}{\bibfnamefont{J.}~\bibnamefont{Fournier}}
  \bibnamefont{et~al.}, \bibinfo{journal}{Physica B}
  \textbf{\bibinfo{volume}{163}}, \bibinfo{pages}{493 } (\bibinfo{year}{1990}).

\bibitem[{\citenamefont{Ichas et~al.}(2001)}]{Ichas01}
\bibinfo{author}{\bibfnamefont{V.}~\bibnamefont{Ichas}} \bibnamefont{et~al.},
  \bibinfo{journal}{Phys. Rev. B} \textbf{\bibinfo{volume}{63}},
  \bibinfo{pages}{045109} (\bibinfo{year}{2001}).

\bibitem[{\citenamefont{Gouder et~al.}(2000)}]{Gouder00}
\bibinfo{author}{\bibfnamefont{T.}~\bibnamefont{Gouder}} \bibnamefont{et~al.},
  \bibinfo{journal}{Phys. Rev. Lett.} \textbf{\bibinfo{volume}{84}},
  \bibinfo{pages}{3378} (\bibinfo{year}{2000}).

\bibitem[{\citenamefont{Havela et~al.}(2002)}]{Havela02}
\bibinfo{author}{\bibfnamefont{L.}~\bibnamefont{Havela}} \bibnamefont{et~al.},
  \bibinfo{journal}{Phys. Rev. B} \textbf{\bibinfo{volume}{65}},
  \bibinfo{pages}{235118} (\bibinfo{year}{2002}).

\bibitem[{\citenamefont{Wachter}(2003)}]{Wachter03}
\bibinfo{author}{\bibfnamefont{P.}~\bibnamefont{Wachter}},
  \bibinfo{journal}{Solid State Commun.} \textbf{\bibinfo{volume}{127}},
  \bibinfo{pages}{599 } (\bibinfo{year}{2003}).

\bibitem[{\citenamefont{Durakiewicz and all}(2004)}]{Durakiewicz04}
\bibinfo{author}{\bibfnamefont{T.}~\bibnamefont{Durakiewicz}} \bibnamefont{and}
  \bibinfo{author}{\bibnamefont{all}}, \bibinfo{journal}{Phys. Rev. B}
  \textbf{\bibinfo{volume}{70}}, \bibinfo{pages}{205103}
  (\bibinfo{year}{2004}).

\bibitem[{\citenamefont{Oppeneer et~al.}(2000)\citenamefont{Oppeneer, Kraft,
  and Brooks}}]{Oppeneer00}
\bibinfo{author}{\bibfnamefont{P.~M.} \bibnamefont{Oppeneer}},
  \bibinfo{author}{\bibfnamefont{T.}~\bibnamefont{Kraft}}, \bibnamefont{and}
  \bibinfo{author}{\bibfnamefont{M.~S.~S.} \bibnamefont{Brooks}},
  \bibinfo{journal}{Phys. Rev. B} \textbf{\bibinfo{volume}{61}},
  \bibinfo{pages}{12825} (\bibinfo{year}{2000}).

\bibitem[{\citenamefont{Shorikov et~al.}(2005)}]{Shorikov05}
\bibinfo{author}{\bibfnamefont{A.~O.} \bibnamefont{Shorikov}}
  \bibnamefont{et~al.}, \bibinfo{journal}{Phys. Rev. B}
  \textbf{\bibinfo{volume}{72}}, \bibinfo{pages}{024458}
  (\bibinfo{year}{2005}).

\bibitem[{\citenamefont{Pourovskii et~al.}(2005)\citenamefont{Pourovskii,
  Katsnelson, and Lichtenstein}}]{Pourovskii05}
\bibinfo{author}{\bibfnamefont{L.~V.} \bibnamefont{Pourovskii}},
  \bibinfo{author}{\bibfnamefont{M.~I.} \bibnamefont{Katsnelson}},
  \bibnamefont{and} \bibinfo{author}{\bibfnamefont{A.~I.}
  \bibnamefont{Lichtenstein}}, \bibinfo{journal}{Phys. Rev. B}
  \textbf{\bibinfo{volume}{72}}, \bibinfo{pages}{115106}
  (\bibinfo{year}{2005}).

\bibitem[{\citenamefont{Suzuki and Oppeneer}(2009)}]{Suzuki09}
\bibinfo{author}{\bibfnamefont{M.~T.} \bibnamefont{Suzuki}} \bibnamefont{and}
  \bibinfo{author}{\bibfnamefont{P.~M.} \bibnamefont{Oppeneer}},
  \bibinfo{journal}{cond-mat/0909.2163}  (\bibinfo{year}{2009}).

\bibitem[{\citenamefont{Svane}(2006)}]{Svane06}
\bibinfo{author}{\bibfnamefont{A.}~\bibnamefont{Svane}},
  \bibinfo{journal}{Solid State Commun.} \textbf{\bibinfo{volume}{140}},
  \bibinfo{pages}{364 } (\bibinfo{year}{2006}).

\bibitem[{\citenamefont{Shick et~al.}(2007)}]{Shick07}
\bibinfo{author}{\bibfnamefont{A.}~\bibnamefont{Shick}} \bibnamefont{et~al.},
  \bibinfo{journal}{Europhys. Lett.} \textbf{\bibinfo{volume}{77}},
  \bibinfo{pages}{17003} (\bibinfo{year}{2007}).

\bibitem[{\citenamefont{Kotliar et~al.}(2006)}]{KotliarRMP}
\bibinfo{author}{\bibfnamefont{G.}~\bibnamefont{Kotliar}} \bibnamefont{et~al.},
  \bibinfo{journal}{Rev. Mod. Phys.} \textbf{\bibinfo{volume}{78}},
  \bibinfo{eid}{865} (\bibinfo{year}{2006}).

\bibitem[{\citenamefont{Held}(2007)}]{Held}
\bibinfo{author}{\bibfnamefont{K.}~\bibnamefont{Held}}, \bibinfo{journal}{Adv.
  Phys.} \textbf{\bibinfo{volume}{56}}, \bibinfo{pages}{829}
  (\bibinfo{year}{2007}).

\bibitem[{\citenamefont{Toropova et~al.}(2007)}]{Toropova07}
\bibinfo{author}{\bibfnamefont{A.}~\bibnamefont{Toropova}}
  \bibnamefont{et~al.}, \bibinfo{journal}{Phys. Rev. B}
  \textbf{\bibinfo{volume}{76}}, \bibinfo{eid}{155126} (\bibinfo{year}{2007}).

\bibitem[{\citenamefont{Savrasov et~al.}(2001)\citenamefont{Savrasov, Kotliar,
  and Abrahams}}]{Savrasov01}
\bibinfo{author}{\bibfnamefont{S.~Y.} \bibnamefont{Savrasov}},
  \bibinfo{author}{\bibfnamefont{G.}~\bibnamefont{Kotliar}}, \bibnamefont{and}
  \bibinfo{author}{\bibfnamefont{E.}~\bibnamefont{Abrahams}},
  \bibinfo{journal}{Nature} \textbf{\bibinfo{volume}{410}},
  \bibinfo{pages}{793} (\bibinfo{year}{2001}).

\bibitem[{\citenamefont{Zhu et~al.}(2007)}]{Zhu07}
\bibinfo{author}{\bibfnamefont{J.-X.} \bibnamefont{Zhu}} \bibnamefont{et~al.},
  \bibinfo{journal}{Phys. Rev. B} \textbf{\bibinfo{volume}{76}},
  \bibinfo{eid}{245118} (\bibinfo{year}{2007}).

\bibitem[{\citenamefont{Shim et~al.}(2007)\citenamefont{Shim, Haule, and
  Kotliar}}]{Shim07}
\bibinfo{author}{\bibfnamefont{J.~H.} \bibnamefont{Shim}},
  \bibinfo{author}{\bibfnamefont{K.}~\bibnamefont{Haule}}, \bibnamefont{and}
  \bibinfo{author}{\bibfnamefont{G.}~\bibnamefont{Kotliar}},
  \bibinfo{journal}{Nature} \textbf{\bibinfo{volume}{446}},
  \bibinfo{pages}{513} (\bibinfo{year}{2007}).

\bibitem[{\citenamefont{Cowan}(1981)}]{Cowan}
\bibinfo{author}{\bibfnamefont{R.~D.} \bibnamefont{Cowan}},
  \emph{\bibinfo{title}{The Theory of Atomic Structure and Spectra}}
  (\bibinfo{publisher}{Univ. California Press, Berkeley},
  \bibinfo{year}{1981}).

\bibitem[{\citenamefont{Coleman}(1984)}]{Coleman84}
\bibinfo{author}{\bibfnamefont{P.}~\bibnamefont{Coleman}},
  \bibinfo{journal}{Phys. Rev. B} \textbf{\bibinfo{volume}{29}},
  \bibinfo{pages}{3035} (\bibinfo{year}{1984}).

\bibitem[{\citenamefont{Kotliar and Ruckenstein}(1986)}]{Kotliar86}
\bibinfo{author}{\bibfnamefont{G.}~\bibnamefont{Kotliar}} \bibnamefont{and}
  \bibinfo{author}{\bibfnamefont{A.~E.} \bibnamefont{Ruckenstein}},
  \bibinfo{journal}{Phys. Rev. Lett.} \textbf{\bibinfo{volume}{57}},
  \bibinfo{pages}{1362} (\bibinfo{year}{1986}).

\bibitem[{\citenamefont{Bickers et~al.}(1987)\citenamefont{Bickers, Cox, and
  Wilkins}}]{Bickers87}
\bibinfo{author}{\bibfnamefont{N.~E.} \bibnamefont{Bickers}},
  \bibinfo{author}{\bibfnamefont{D.~L.} \bibnamefont{Cox}}, \bibnamefont{and}
  \bibinfo{author}{\bibfnamefont{J.~W.} \bibnamefont{Wilkins}},
  \bibinfo{journal}{Phys. Rev. B} \textbf{\bibinfo{volume}{36}},
  \bibinfo{pages}{2036} (\bibinfo{year}{1987}).

\bibitem[{\citenamefont{Kroha et~al.}(2003)}]{Kroha03}
\bibinfo{author}{\bibfnamefont{J.}~\bibnamefont{Kroha}} \bibnamefont{et~al.},
  \bibinfo{journal}{Physica E} \textbf{\bibinfo{volume}{18}},
  \bibinfo{pages}{69 } (\bibinfo{year}{2003}).

\end{thebibliography}

\end{document}